\documentclass[final]{aipproc}\layoutstyle{6x9}\begin{document}
\title{Generalized Instantaneous Bethe--Salpeter Equation and Exact
Quark Propagators}

\author{Wolfgang Lucha}{address={Institute for High Energy Physics,
Austrian Academy of Sciences,\\Nikolsdorfergasse 18, A-1050
Vienna, Austria\\E-mail: wolfgang.lucha@oeaw.ac.at}}
\author{Franz F.~Sch\"oberl}{address={Department of Theoretical
Physics, University of Vienna,\\Boltzmanngasse 5, A-1090 Vienna,
Austria\\E-mail: franz.schoeberl@univie.ac.at}}\maketitle

Assuming all interactions to be instantaneous \emph{and} the exact
fermion propagator $S$ to be approximately given by ${\rm
i}\,S^{-1}(p)=A(\mathbf{p}^2)\,p\hspace{-1.07ex}/-B(\mathbf{p}^2),$
i.e., with its scalar functions $A$ and $B$ depending only on
$\mathbf{p},$ a three-dimensional reduction of the homogeneous
Bethe--Salpeter equation retaining, in contrast to the Salpeter
equation, the exact propagators (crucial for, e.g., a proper
incorporation of dynamical chiral symmetry breakdown) is
constructed \cite{Lucha05:IBSEWEP}. For spherically symmetric
interactions the resulting bound-state equation reduces to a set
of (coupled) integral equations for the independent radial
components of the bound-state amplitude \cite{RedEq}, which may be
solved by conversion into an equivalent matrix problem
\cite{Lucha00/01:IBSE}. Adopting as light-quark propagator the
solution of the quark Dyson--Schwinger equation found within a
model \cite{Maris} consistent with the QCD axial-vector
Ward--Takahashi identity, a tentative application
\cite{Lucha05:EQPIBSE} to ``pion-like'' pseudoscalar
($J^{PC}=0^{-+}$) mesons considered as quark--antiquark bound
states formed by linear confining interactions of time-component
Lorentz-vector Dirac structure $\gamma^0\otimes\gamma^0$ indicates
that our ``exact-propagator instantaneous formalism'' yields
significantly smaller spacings of its bound-state mass eigenvalues
than those obtained from Salpeter's equation for ``reasonable''
constituent light-quark masses.

\bibliographystyle{aipproc}\end{document}